\documentclass[11pt,a4paper]{article}

\usepackage{latexsym,amssymb, amsthm}
\usepackage[centertags]{amsmath}
\usepackage{epsfig,pifont,rotating}
\usepackage[inactive]{srcltx}  

\usepackage{graphics,helvet,times,mathptm,epic,eepic}

\usepackage{color}

\usepackage{newlfont}

\usepackage{amsfonts,bbm}

\newenvironment{keywords}{ \noindent {\small\bf Key Words}:}{ }

\def\bd{\begin{description}}
\def\ed{\end{description}}

\def\beq{\begin{equation}}
\def\eeq{\end{equation}}
\def\bea{\begin{eqnarray}}
\def\eea{\end{eqnarray}}
\def\beas{\begin{eqnarray*}}
\def\eeas{\end{eqnarray*}}

\newtheorem{theorem}{Theorem}[section]

\newtheorem{corollary}{Corollary}[section]

\theoremstyle{remark}

\newtheorem{example}{Example}[section]

\newcommand{\M}{\mathcal{M}}
\newcommand{\abs}[1]{\left\vert#1\right\vert}

\newcommand{\seq}[1]{\left<#1\right>}

\begin{document}

\title{\textbf{\textsc{Observability of Turing Machines:\\ a Refinement of the Theory of
Computation}}}

\newcommand{\nms}{\normalsize}
\author{  {   \bf Yaroslav D. Sergeyev\footnote{Yaroslav D. Sergeyev, Ph.D.,
 D.Sc., is Distinguished Profesor at the University of Calabria, Rende, Italy.
 He is also Full Professor (a part-time contract) at the N.I.~Lobatchevsky State University,
  Nizhni Novgorod, Russia and Affiliated Researcher at the Institute of High Performance
  Computing and Networking of the National Research Council of Italy. {\tt  yaro@si.deis.unical.it}}
  \hspace{2mm}and Alfredo Garro\footnote{Alfredo Garro, Ph.D., is Assistant
Professor at the University of Calabria, Rende, Italy.  {\tt
garro@si.deis.unical.it }} \footnote{The authors thank the
anonymous reviewers for their useful suggestions. This research
was partially supported by the Russian Federal Program
``Scientists and Educators in Russia of Innovations", contract
number 02.740.11.5018.} }\\ \\ [-2pt]
      \nms Dipartimento di Elettronica, Informatica e Sistemistica,\\[-4pt]
       \nms   Universit\`a della Calabria,\\[-4pt]
       \nms 87030 Rende (CS)  -- Italy\\ \\[-4pt]
               }

\date{}

\maketitle

\begin{abstract}
The Turing machine is one of the   simple abstract computational
devices that can be used  to   investigate the   limits of
computability. In this paper, they are considered from several
points   of view that emphasize the importance and the relativity
of mathematical languages used to describe the Turing machines. A
deep investigation is performed on the interrelations  between
mechanical computations and their mathematical descriptions
emerging when a human (the researcher) starts to describe a Turing
machine (the object of the study) by different mathematical
languages (the instruments of investigation). Together with
traditional mathematical languages using such concepts as
`enumerable sets' and `continuum' a new computational methodology
allowing one   to measure the number of elements of different
infinite sets is used in this paper. It is shown how mathematical
languages used to describe the machines limit our possibilities to
observe them. In particular, notions of   observable deterministic
and non-deterministic Turing machines are introduced and
conditions ensuring that the latter  can be simulated by the
former  are established.
 \end{abstract}

\begin{keywords}
Theory of automatic computations, observability of Turing
machines, relativity of mathematical languages, infinite sets,
Sapir–-Whorf thesis.
 \end{keywords}

\newpage
\section{Introduction}
\label{s0_Turing}

The fundamental nature of the concept \textit{automatic
computations} attracted a great attention of mathematicians (and
later  of computer scientists) since 1930's (see
\cite{Church,Kleene,Kolmogorov,Kolmogorov_Uspensky,Markov,Mayberry,Post,Turing}
and more recent monographs
\cite{Ausiello,Barry_Cooper,Davis,Hopcroft_Ullman}). At that time,
this strong impetus for understanding what is computable was
actively supported by David Hilbert who believed that all of
Mathematics could be precisely axiomatized. Several mathematicians
from around the world proposed their independent definitions of
what it means to be computable and what it means an automatic
computing machine. In order to perform a rigorous study of
sequential computations, they worked with different mathematical
models of computing machines. Surprisingly, it has been discovered
(see detailed discussions on this topic in, e.g.,
\cite{Ausiello,Barry_Cooper,Davis}) that all of these models were
equivalent, e.g., anything computable in the $\lambda$-calculus is
computable by a Turing machine.

 In spite of the fact that the famous
results of Church, G\"odel, and Turing have shown that Hilbert's
programme cannot be realized, the idea of finding an adequate set
of axioms for one or another field of  Mathematics continues to be
among the most attractive goals for contemporary mathematicians as
well. Usually,   when it is necessary  to define a concept or an
object, logicians   try to introduce a number of axioms describing
the object. However, this way is fraught with danger because of
the following reasons.

First, when we describe a mathematical object or concept we are
limited by the expressive capacity of the language we use to make
this description. A   richer language allows us to say more about
the object and a weaker language -- less. Thus, development of the
mathematical (and not only mathematical) languages leads to a
continuous necessity of a transcription and specification of
axiomatic systems. Second, there is no   guarantee that the chosen
axiomatic system defines `sufficiently well' the required concept
and a continuous comparison with practice is required in order to
check the goodness of the accepted set of axioms. However, there
cannot be again any guarantee that the new version will be the
last and definitive one. Finally, the third limitation already
mentioned above    has been discovered by G\"odel in his two
famous incompleteness theorems (see~\cite{Godel_1931}).

In linguistics, the relativity of the language with respect to the
world around us has been formulated in the form of the
Sapir–-Whorf thesis (see~\cite{Whorf,Gumperz_Levinson,Lucy,Sapir})
also known as the `linguistic relativity  thesis' (that has also
interesting relations to the ideas of K.E.~Iverson exposed in his
Turing lecture \cite{Iverson}). As becomes  clear from  its name,
the thesis does not accept the idea of the universality of
language and  postulates that the nature of a particular language
influences the thought of its speakers. The thesis challenges the
possibility of perfectly representing the world with language,
because it implies that the mechanisms of any language condition
the thoughts of its speakers.

In this paper, we study the relativity   of mathematical languages
in situations where they are used to observe and to  describe
automatic computations (we consider the traditional computational
paradigm  mainly following results of Turing (see \cite{Turing})
whereas emerging computational paradigms (see, e.g.
~\cite{Adamatzky,Nielsen,Walster_1,Zilinskas_informatica}) are not
considered here). Let us illustrate the concept of the relativity
of mathematical languages by the following example. In his study
published in \textit{Science} (see \cite{Gordon}), Peter Gordon
describes a primitive tribe living in Amazonia -- Pirah\~{a} --
that uses a very simple numeral system\footnote{We remind that
\textit{numeral}  is a symbol or group of symbols that represents
a \textit{number}. The difference between numerals and numbers is
the same as the difference between words and the things they refer
to. A \textit{number} is a concept that a \textit{numeral}
expresses. The same number can be represented by different
numerals. For example, the symbols `9', `nine', and `IX' are
different numerals, but they all represent the same number.} for
counting: one, two, `many'. For Pirah\~{a}, all quantities larger
than two are just `many' and such operations as 2+2 and 2+1 give
the same result, i.e., `many'. By using their weak numeral system
Pirah\~{a} are not able to see, for instance, numbers 3, 4, and 5,
to execute arithmetical operations with them, and, in general, to
say anything about these numbers because in their language there
are neither words nor concepts for that.

The numeral system of Pirah\~{a} has another interesting feature
particularly interesting in the context of  the study presented in
this paper: \beq
 \mbox{`many'}+ 1= \mbox{`many'},   \hspace{3mm}
\mbox{`many'} + 2 = \mbox{`many'}, \hspace{3mm} \mbox{`many'}+
\mbox{`many'} = \mbox{`many'}.
 \label{Turing_1}
 \eeq
 These relations are very
familiar to us  in the context of our views on infinity used in
the   calculus
 \beq
  \infty + 1= \infty,    \hspace{1cm}    \infty
+ 2 = \infty, \hspace{1cm}    \infty + \infty = \infty.
 \label{Turing_2}
 \eeq
Thus, the modern mathematical numeral systems allow us to
distinguish a larger quantity of finite numbers with respect to
Pirah\~{a} but give  similar results  when we speak  about
infinite numbers. Formulae (\ref{Turing_1}) and (\ref{Turing_2})
lead us to the following observation: \textit{Probably our
difficulty in working with infinity is not connected to the nature
of infinity but is a result of inadequate numeral systems used to
express infinite numbers.} Analogously, Pirah\~{a} do not
distinguish numbers 3 and 4 not due to the  nature of these
numbers but due to the weakness of their numeral system.

This remark is   important with respect to the computability
context because of the following reason. Investigations of
traditional computational models (we do not discuss emerging
computational paradigms, see, e.g. \cite{Cantor}) executed so far
used for studying infinite computational processes mathematical
instruments developed by Georg Cantor (see \cite{Cantor}) who has
shown that there exist infinite sets having different number of
elements. In the theory of computations, two infinite sets --
countable sets and continuum -- are used mainly. Cantor has
proved, by using his famous diagonal argument, that the
cardinality, $\aleph_0$, of the set, $\mathbb{N}$, of natural
numbers is less than the cardinality, $C$, of real numbers    $x
\in[0,1]$.

Cantor  has also developed an arithmetic for the infinite cardinal
numbers. Some of the operations of this arithmetic including
$\aleph_0$ and $C$ are given below:
\[
\aleph_0 + 1 \,\,\,\, =     \aleph_0,  \hspace{1cm} \aleph_0 + 2
\,\,\,\, =     \aleph_0,  \hspace{1cm}  \aleph_0 + \aleph_0
\,\,\,\,  =   \aleph_0,
\]
\[
C + 1 \,\,\,\, =     C, \hspace{1cm} C + 2 \,\,\,\, = C,
\hspace{1cm}  C + C \,\,\,\,  = C, \hspace{1cm} C  + \aleph_0
\,\,\,\, = C.
\]
Again, it is possible to see a clear similarity with the
arithmetic operations used in the numeral system of Pirah\~{a}.

   Advanced
contemporary numeral systems  enable us to distinguish within
`many'   various large finite numbers. As a result, we can use
large finite numbers in computations and  construct mathematical
models involving them. Analogously, if we were be able to
distinguish more infinite numbers probably we could  understand
better the nature of the sequential automatic computations (remind
the famous phrase of Ludwig Wittgenstein: `The limits of my
language are the limits of my world.').

The goal of this paper is to study Turing machines using a new
approach introduced in \cite{Sergeyev,www,informatica} and
allowing one to write down different finite, infinite, and
infinitesimal numbers by a finite number of symbols as particular
cases of a unique framework. Its applications in several fields
can be found in \cite{Sergeyev,Menger,Korea,Dif_Calculus,first}.
It is worthy to mention also that the new computational
methodology has given a possibility to introduce the Infinity
Computer (see \cite{www} and the European patent
\cite{Sergeyev_patent}) working numerically with finite, infinite,
and infinitesimal numbers (its software simulator has already been
realized).

The rest of the paper is structured as follows.   In
Section~\ref{s1_Turing},   a brief   introduction to  the new
methodology is given.  Due to a rather unconventional character of
the new methodology, the authors kindly recommend the reader to
study the survey \cite{informatica} (downloadable from \cite{www})
before approaching Sections~\ref{s2_Turing}~--~\ref{s4_Turing}.

Section~\ref{s2_Turing} presents some preliminary results
regarding description of infinite sequences by using a new numeral
system. Section~\ref{s3_Turing} shows that the introduced
methodology applied  together with a new numeral system allows one
to have a fresh look at mathematical descriptions of Turing
machines. A deep investigation is performed on the  interrelations
between mechanical computations and their mathematical
descriptions emerging when a human (the researcher) starts to
describe a Turing machine (the object of the study) by different
mathematical languages (the instruments of investigation).
Mathematical descriptions of automatic computations obtained
  by using the traditional language and the new one are compared
and discussed. An example of the comparative usage of both
languages is given in Section~\ref{s4_Turing} where  they are
applied  for descriptions
 of deterministic and non-deterministic Turing machines.
After all, Section~\ref{s5_Turing} concludes the paper.

\section{Methodology and a new numeral system}
\label{s1_Turing}

In this section, we give just a brief introduction to the
methodology of the new approach \cite{Sergeyev,informatica}
dwelling only on the issues directly related to the subject of the
paper.  This methodology will be used in the subsequent sections
to study Turing machines and to obtain some more detailed  results
related to the further understanding of what is effectively
computable -- the problem that was stated and widely discussed in
\cite{Church,Turing}.

We start by introducing three postulates that will fix our
methodological positions (having a strong applied character) with
respect to infinite and infinitesimal quantities and Mathematics,
in general.

\textbf{Postulate 1.} \textit{There exist infinite and
infinitesimal objects but   human beings and machines are able to
execute only a finite number of operations.}

\textbf{Postulate 2.} \textit{We shall not   tell \textbf{what
are} the mathematical objects we deal with; we just shall
construct more powerful tools that will allow us to improve our
capabilities to observe and to describe properties of mathematical
objects.}

\textbf{Postulate 3.} \textit{The principle `The part is less than
the whole' is applied to all numbers (finite, infinite, and
infinitesimal) and to all sets and processes (finite and
infinite).}

In Physics,   researchers use tools to describe the object of
their study and the used instrument influences results of
observations and restricts possibilities of observation of the
object. Thus, there exists the philosophical triad -- researcher,
object of investigation, and tools used to observe the object.
 Postulates 1--3   emphasize  existence of this
triad in Mathematics and Computer Science, as well. Mathematical
languages (in particular, numeral systems) are among the tools
used by mathematicians to observe and to describe mathematical
objects. As a consequence,  very often difficulties that we find
solving mathematical problems are related not to their nature but
to inadequate mathematical languages used to solve them.

It is necessary to notice that due to the declared applied
statement fixed by Postulates 1--3, such concepts as bijection,
numerable and continuum sets, cardinal and ordinal numbers cannot
be used in this paper because they belong to the theories working
with different assumptions. As a consequence, the new approach is
different also with respect to the non-standard analysis
introduced in \cite{Robinson} and built using Cantor's ideas.
However, the approach used here does not contradict Cantor. In
contrast, it evolves his deep ideas regarding existence of
different infinite numbers in a more applied way and can be viewed
as a more strong lens of our mathematical microscope that allows
one, e.g., not only to separate different classes of infinite sets
but also to measure the number of elements of some infinite sets.

By accepting Postulate~1   we admit that it is not possible to
have  a complete description of infinite processes and sets due to
our finite capabilities.  For instance,   we accept that we are
not able to observe all elements of an infinite set (this issue
will be discussed in detail hereinafter).

It is important to emphasize that our point of view on axiomatic
systems is also more applied than the traditional one. Due to
Postulate~2,    mathematical objects are not defined by axiomatic
systems that just determine formal rules for operating with
certain numerals reflecting some properties of the studied
mathematical objects.

Due to Postulate 3, infinite and infinitesimal numbers should be
managed in the same manner as we are used to deal with finite
ones. This Postulate in our opinion very well reflects
organization of the world around us but   in many traditional
infinity theories   it is true only for finite numbers. Due to
Postulate~3, the traditional point of view on infinity accepting
such results as $\infty + 1= \infty$ should be substituted in   a
way ensuring that $\infty + 1 > \infty$.

This methodological program has been realized in
\cite{Sergeyev,informatica} where a new powerful numeral system
has been developed. This system gives   a possibility to execute
\textit{numerical} computations not only with finite numbers but
also with infinite and infinitesimal ones in accordance with
Postulates 1--3. The main idea consists of measuring  infinite and
infinitesimal quantities   by different (infinite, finite, and
infinitesimal) units of measure.

A new infinite unit of measure   has been introduced for this
purpose in \cite{Sergeyev,informatica} in accordance with
Postulates 1--3 as the number of elements of the set $\mathbb{N}$
of natural numbers. It is expressed by a new numeral \ding{172}
called \textit{grossone}.

It is necessary to emphasize immediately that the infinite number
\ding{172} is not either Cantor's $\aleph_0$ or $\omega$.
Particularly, it has both cardinal and ordinal properties as usual
finite natural numbers. Formally, grossone is introduced as a new
number by describing its properties postulated by the
\textit{Infinite Unit Axiom}   (see \cite{Sergeyev,informatica}).
This axiom is added to axioms for real numbers similarly to
addition of the axiom determining zero to axioms of natural
numbers when integer numbers are introduced. Again, we speak about
axioms of real numbers in sense of Postulate~2, i.e., axioms
define formal rules of operations with numerals in a given numeral
system.

Inasmuch as it has been postulated that grossone is a number,  all
other axioms for numbers hold for it, too. Particularly,
associative and commutative properties of multiplication and
addition, distributive property of multiplication over addition,
existence of   inverse  elements with respect to addition and
multiplication hold for grossone as for finite numbers. This
means, for example, that  the following relations hold for
grossone, as for any other number
 \beq
 0 \cdot \mbox{\ding{172}} =
\mbox{\ding{172}} \cdot 0 = 0, \hspace{3mm}
\mbox{\ding{172}}-\mbox{\ding{172}}= 0,\hspace{3mm}
\frac{\mbox{\ding{172}}}{\mbox{\ding{172}}}=1, \hspace{3mm}
\mbox{\ding{172}}^0=1, \hspace{3mm}
1^{\mbox{\tiny{\ding{172}}}}=1, \hspace{3mm}
0^{\mbox{\tiny{\ding{172}}}}=0.
 \label{3.2.1}
       \eeq

Let us comment upon the nature of grossone by some illustrative
examples (see the survey \cite{informatica} for a detailed
discussion).

\begin{example}
\label{e1}  Infinite numbers constructed using grossone  can be
interpreted in terms of the number of elements of infinite sets.
For example, $\mbox{\ding{172}}-2$ is the number of elements of a
set $B=\mathbb{N}\backslash\{b_1, b_2\}$ where $b_1, b_2 \in
\mathbb{N}$. Analogously, $\mbox{\ding{172}}+1$ is the number of
elements of a set $A=\mathbb{N}\cup\{a\}$, where $a \notin
\mathbb{N}$. Due to Postulate~3, integer positive numbers that are
larger than grossone do not belong to $\mathbb{N}$ but also can be
easily interpreted. For instance,   $\mbox{\ding{172}}^3$ is the
number of elements of the set
  $V$, where
\[
 \hspace{27mm} V  =
\{ (a_1, a_2, a_3)  : a_1 \in   \mathbb{N}, a_2 \in   \mathbb{N},
a_3 \in   \mathbb{N} \}.
     \hspace{3cm} \Box
 \]
\end{example}

\begin{example}
\label{e2}  Grossone has been introduced as the quantity of
natural numbers. Similarly  to the set
 \beq
  A=\{1, 2, 3, 4, 5\}
\label{4.1.deriva_0}
 \eeq
   having
5 elements where 5 is the largest number in $A$, \ding{172} is the
largest \textit{infinite} natural number\footnote{This fact is one
of the important methodological differences with respect to
non-standard analysis theories where it is supposed that infinite
numbers do not belong to $\mathbb{N}$.} and $\mbox{\ding{172}} \in
\mathbb{N}$. As a consequence, the set, $\mathbb{N}$, of natural
numbers can be written  in the form
 \beq
\mathbb{N} = \{ 1,2,3, \hspace{5mm} \ldots  \hspace{5mm}
\mbox{\ding{172}}-2, \hspace{2mm}\mbox{\ding{172}}-1, \hspace{2mm}
\mbox{\ding{172}} \}.   \label{4.1}
       \eeq
Traditional numeral systems did not allow us to see infinite
natural numbers $\ldots    \mbox{\ding{172}}-2,
\mbox{\ding{172}}-1,$ $ \mbox{\ding{172}}$. Similarly, the
Pirah\~{a} are not able to see finite natural numbers greater
than~2. In spite of this fact, these numbers (e.g., 3 and 4)
belong to $\mathbb{N}$ and are visible if one uses a more powerful
numeral system. Thus, we have the same object of observation --
the set $\mathbb{N}$ -- that can be observed by different
instruments -- numeral systems -- with different accuracies (see
Postulate~2). \hfill $\Box$
\end{example}

As it has been mentioned above, the introduction of  the numeral
$\mbox{\ding{172}}$ allows us to introduce various numerals that
can be used to express integer positive numbers larger than
grossone such as $\mbox{\ding{172}}^2$, $\mbox{\ding{172}}^3-4$,
and also $2^{\mbox{\tiny{\ding{172}}}},
10^{\mbox{\tiny{\ding{172}}}}+3$, etc. (their meaning will be
explained soon). This leads us to the necessity to introduce the
set of \textit{extended natural numbers} (including $\mathbb{N}$
as a proper subset) indicated as $\widehat{\mathbb{N}}$ where
 \beq
  \widehat{\mathbb{N}} = \{
1,2, \ldots ,\mbox{\ding{172}}-1, \mbox{\ding{172}},
\mbox{\ding{172}}+1, \mbox{\ding{172}}+2, \mbox{\ding{172}}+3,
\ldots , \mbox{\ding{172}}^2-1, \mbox{\ding{172}}^2,
\mbox{\ding{172}}^2+1, \ldots \}. \label{4.2.2}
       \eeq

It is useful  to notice that,  due to Postulates 1 and 2, the new
numeral system   cannot give answers to \textit{all} questions
regarding infinite sets. A mathematical language can allow one to
formulate a question but not its answer. For instance, it is
possible to formulate the question: `What is the number of
elements of the set  $\widehat{\mathbb{N}}$?' but the answer to
this question cannot be expressed within a numeral system using
only \ding{172}. It is necessary to introduce in a reasonable way
a more powerful numeral system by defining new numerals (for
instance, \ding{173}, \ding{174}, etc.).

\begin{example}
\label{e1_Turing} Let us consider the set of even numbers,
$\mathbb{E}$, from the traditional point of view. Cantor's
approach establishes the following one-to-one correspondence with
the set of all natural numbers, $\mathbb{N}$, in spite of the fact
that $\mathbb{E}$ is a part of $\mathbb{N}$:
 \beq
\begin{array}{lccccccc}
  \mbox{even numbers:}   & \hspace{5mm} 2, & 4, & 6, & 8,  & 10, & 12, & \ldots    \\

& \hspace{5mm} \updownarrow &  \updownarrow & \updownarrow  & \updownarrow  & \updownarrow  &  \updownarrow &   \\

  \mbox{natural numbers:}& \hspace{5mm}1, &  2, & 3, & 4 & 5,
       & 6,  &    \ldots \\
     \end{array}
\label{4.4.1}
 \eeq
This result can be viewed in the following way: traditional
mathematical tools do not allow us to distinguish inside the class
of enumerable sets infinite sets having different number of
elements.

From the new point of view, the one-to-one correspondence cannot
be used as a \textit{tool} for working with infinite sets because,
due to Postulate~1,  we are able to execute only a finite number
of operations and the sets $\mathbb{E}$ and $\mathbb{N}$ are
infinite. However, analogously to (\ref{4.1}), the set,
$\mathbb{E}$, of even natural numbers can be written (see
\cite{informatica} for a detailed discussion) in the form
 \beq
\mathbb{E} = \{ 2,4,6 \hspace{5mm} \ldots  \hspace{5mm}
\mbox{\ding{172}}-4, \hspace{2mm}\mbox{\ding{172}}-2, \hspace{2mm}
\mbox{\ding{172}} \},   \label{4.1.0}
       \eeq
since \ding{172} is even  and the number of elements of the set of
even natural numbers is equal to $\frac{\mbox{\ding{172}}}{2}$.
Note that the next even number is $\mbox{\ding{172}}+2$ but it is
not natural because $\mbox{\ding{172}}+2  > \mbox{\ding{172}}$
(see (\ref{4.2.2})), it is extended natural. Thus, we can write
down not only initial (as it is  done traditionally) but also the
final part of (\ref{4.4.1}) as follows
  \beq
\begin{array}{cccccccccc}
 2, & 4, & 6, & 8,  & 10, & 12, & \ldots  &
\mbox{\ding{172}} -4,  &    \mbox{\ding{172}}  -2,   &    \mbox{\ding{172}}    \\
 \updownarrow &  \updownarrow & \updownarrow  &
\updownarrow  & \updownarrow  &  \updownarrow  & &
  \updownarrow    & \updownarrow   &
  \updownarrow
   \\
 1, &  2, & 3, & 4 & 5, & 6,   &   \ldots  &    \frac{\mbox{\ding{172}}}{2} - 2,  &
     \frac{\mbox{\ding{172}}}{2} - 1,  &    \frac{\mbox{\ding{172}}}{2}   \\
     \end{array}
\label{4.4.2}
 \eeq
concluding so (\ref{4.4.1})   in a complete accordance with
Postulate~3.
 \hfill $\Box$
\end{example}

 Note that record (\ref{4.4.2}) does not affirm  that
we have established the one-to-one correspondence among
\textit{all}  even numbers and a half of natural ones. We cannot
do this  because, due to Postulate~1, we can execute only a finite
number of operations and the considered sets are infinite. The
symbols `$\ldots$' in (\ref{4.4.2}) indicate that there are
infinitely many numbers between $12$ and $\mbox{\ding{172}} -4$ in
the first line and between $6$ and $\frac{\mbox{\ding{172}}}{2} -
2$ in the second line.  The record (\ref{4.4.2}) affirms that for
any even natural number expressible in the chosen numeral system
it is possible to indicate the corresponding natural number in the
lower row of (\ref{4.4.2}) if it is also expressible in this
numeral system.

We conclude the discussion upon   Example~\ref{e1_Turing}  by the
following remark. With respect to our methodology, the
mathematical results obtained by Cantor in (\ref{4.4.1}) and our
results (\ref{4.4.2}) do not contradict  each other. \textit{They
both are correct with respect to mathematical languages used to
express them.} This relativity is very important and it has been
emphasized in Postulate~2. The result (\ref{4.4.1}) is correct in
Cantor's language and the more powerful language developed in
\cite{Sergeyev,informatica} allows us to obtain a more precise
result (\ref{4.4.2}) that is correct in the new language.

The choice of the mathematical language  depends on the practical
problem that is to be   solved and on the accuracy required for
such a solution.   The result (\ref{4.4.1}) just means that
Cantor's mathematical tools do not allow one to distinguish two
observed mathematical objects, namely, the number of elements of
the sets $\mathbb{E}$ and $\mathbb{N}$ from the point of view of
the number of their elements. If one is satisfied with this
accuracy, this answer can be used (and \textit{was used} since
Cantor has published his results in 1870's) in practice.

However, if one needs a more precise result, it is necessary to
introduce a more powerful mathematical language (a numeral system
in this case) allowing one to express the required answer in a
more accurate way. Obviously, it is not possible to mix languages.
For instance, the question `what is the result of the operation
`many'+4?', where `many' belongs to the numeral system of
Pirah\~{a}, is nonsense.

\section{Infinite sequences} \label{s2_Turing}

In the traditional definition of the Turing machine the notion of
infinity is   used in a strong form (see \cite{Turing} and, e.g.,
\cite{Davis}). First, the Turing machine has an infinite
one-dimensional tape divided into cells and its outputs are
computable (infinite) sequences of numerals. Second, an infinite
sequence of operations can be executed by the machine and it is
supposed the availability of an infinite time for the computation.
Turing writes in pages~232 and~233 of \cite{Turing}:

\begin{quote}
 \textit{\textbf{Computing machines. }}\\
 If an $a$-machine prints two kinds of symbols,
of which the first kind (called figures) consists entirely of 0
and 1 (the others being called symbols of the second kind), then
the machine will be called a computing machine. If the machine is
supplied with a blank tape and set in motion, starting from the
correct initial $m$-configuration, the subsequence of the symbols
printed by it which are of the first kind will be called the
\textit{sequence computed by the machine}. The real number whose
expression as a binary decimal is obtained by prefacing this
sequence by a decimal point is called the \textit{number computed
by the machine}. [...]

\textit{\textbf{Circular and circle-free machines. }}\\
 If
a computing machine never writes down more than a finite number of
symbols of the first kind it will be called \textit{circular}.
Otherwise it is said to be \textit{circle-free}.  [...]

\textit{\textbf{Computable sequences and numbers.}}\\
A sequence is said to be computable if it can be computed by a
circle-free machine.
\end{quote}

It  is clear  that the notion of the infinite sequence becomes
very important for our study of the Turing machine. Thus, before
considering the notion of the Turing machine from the point of
view of the new methodology, let us explain how the notion of the
infinite sequence can be viewed from the new positions.

Traditionally, an \textit{infinite sequence} $\{a_n\}, a_n \in A,$
$n \in \mathbb{N},$ is defined as a function having the set of
natural numbers, $\mathbb{N}$, as the domain  and a set $A$ as the
codomain. A \textit{subsequence} $\{b_n\}$ is defined as  a
sequence $\{a_n\}$ from which some of its elements have been
removed. In spite of the fact of the removal of the elements from
$\{a_n\}$, the traditional approach does not allow one to
register, in the case where the obtained subsequence $\{b_n\}$ is
infinite,  the fact that $\{b_n\}$ has less elements than the
original infinite sequence $\{a_n\}$.

From the point of view of the new methodology, an infinite
sequence can be considered in a dual way: either as an object of a
mathematical study or as a mathematical instrument developed by
human beings to observe other objects and processes (see
Postulate~2).  First, let us consider it as a mathematical object
and show that the definition of infinite sequences should be done
more precise within the new methodology. The following result (see
\cite{Sergeyev,informatica}) holds. We reproduce here its proof
for the sake of completeness.
\begin{theorem}
\label{t2} The number of elements of any infinite sequence is less
or equal to~\ding{172}.
\end{theorem}

\textit{Proof.}  The new numeral system allows us to express the
number of elements of the set $\mathbb{N}$  as \ding{172}. Thus,
due to the sequence definition given above, any sequence having
$\mathbb{N}$ as the domain  has \ding{172} elements.

The notion of subsequence is introduced as a sequence from which
some of its elements have been removed. Due to Postulate~3, this
means that the resulting subsequence will have less elements than
the original sequence. Thus, we obtain infinite sequences having
the number of members less than grossone.  \hfill $\Box$

It becomes appropriate now to define the \textit{complete
sequence} as an infinite sequence  containing \ding{172} elements.
For example, the sequence   of natural numbers  is complete, the
sequences of even  and odd natural numbers  are not complete
because  they have $\frac{\mbox{\ding{172}}}{2}$ elements each
(see \cite{Sergeyev,informatica}). Thus, the new approach imposes
a more precise description of infinite sequences than the
traditional one.

To define a sequence $\{a_n\}$ in the new language, it is not
sufficient just to give a formula for~$a_n$, we should determine
(as it happens for sequences having a finite number of elements)
its number of elements and/or the first and the last elements of
the sequence. If the number of the first element is equal to one,
we can use the record $\{a_n: k \}$ where $a_n$ is, as usual, the
general element of the sequence and $k$ is the number (that can be
finite or infinite) of members of the sequence.

In connection with this definition the following   question arises
inevitably. Suppose that we have two sequences, for example,
$\{b_n: k_1\}$ and $\{c_n: k_2\}$, where both $k_1$ and $k_2$ are
infinite numbers such that $ k_1 < \mbox{\ding{172}}$ and $ k_2 <
\mbox{\ding{172}}$ but $ k_1 + k_2 > \mbox{\ding{172}}$. Can we
create a new sequence, $\{d_n:k\}$, composed from both of them,
for instance, as it is shown below
 \[
b_1,\hspace{1mm} b_2,\hspace{1mm} \ldots \hspace{1mm}
b_{k_1-2},\hspace{1mm} b_{k_1-1},\hspace{1mm} b_{k_1},\hspace{1mm}
c_1,\hspace{1mm} c_2,\hspace{1mm} \ldots \hspace{1mm}
c_{k_2-2},\hspace{1mm} c_{k_2-1},\hspace{1mm} c_{k_2}
 \]
and which will be the value of the number of its elements $k$?

The answer to this question is `no' because  due to Theorem~
\ref{t2}, a sequence   cannot have more than $\mbox{\ding{172}}$
elements. Thus, the longest sequence is
$\{d_n:\mbox{\ding{172}}\}$. After arriving to the last element
$d_{\mbox{\tiny{\ding{172}}}}$, the sequence
$\{d_n:\mbox{\ding{172}} \}$ will stop. However, the second
sequence can then be started.

\begin{example}
\label{e2_Turing} Suppose that $k_1=\frac{2\mbox{\ding{172}}}{5}$
and $k_2=\frac{4\mbox{\ding{172}}}{5}$.  Then starting from the
element $b_1$ we can arrive at maximum to the element
$c_{\frac{3\mbox{\tiny{\ding{172}}}}{5}}$  being the element $
d_{\mbox{\tiny{\ding{172}}}}$ in the sequence
$\{d_n:\mbox{\ding{172}}\}$ which we   construct. Therefore,
$k=\mbox{\ding{172}}$ and
\[
\underbrace{b_1,\hspace{1mm}  \ldots \hspace{1mm}
b_{\frac{2\mbox{\tiny{\ding{172}}}}{5}},\hspace{1mm}
 c_1,\hspace{1mm}  \ldots
c_{\frac{3\mbox{\tiny{\ding{172}}}}{5}}}_{\mbox{\ding{172}
elements}}, \hspace{1mm}
\underbrace{c_{\frac{3\mbox{\tiny{\ding{172}}}}{5}+1}, \ldots
\hspace{1mm}
c_{\frac{4\mbox{\tiny{\ding{172}}}}{5}}}_{\frac{\mbox{\tiny{\ding{172}}}}{5}
 \mbox{ elements }}.
\]
The remaining members $c_{\frac{3\mbox{\tiny{\ding{172}}}}{5}+1},
\ldots \hspace{1mm} c_{\frac{4\mbox{\tiny{\ding{172}}}}{5}}$ of
the sequence $\{c_n:\frac{4\mbox{\ding{172}}}{5}\}$ will form the
second sequence, $\{g_n: l \}$ having $l=
\frac{4\mbox{\ding{172}}}{5}-\frac{3\mbox{\ding{172}}}{5} =
\frac{\mbox{\ding{172}}}{5}$  elements. Thus, we have formed two
sequences, the first of them is complete and the second is
not.\hfill
 $\Box$
\end{example}

We have already seen the influence  of Postulates~2 and~3 on the
notion of the infinite sequence. Let us study now what Postulate~1
gives us in this context. First, since the object of the study --
the sequence -- has an infinite number of members, it follows from
Postulate~1 that we cannot observe all of them. We can observe
only a finite number of its elements, precisely, those members of
the sequence for which there exist the corresponding numerals in
the chosen numeral system.

\begin{example} \label{e3_Turing}  Let us consider the  numeral system,
$\mathcal{P}$, of Pirah\~{a} able to express only numbers 1 and 2.
If we add to $\mathcal{P}$ the new numeral \ding{172}, we obtain a
new numeral system (we call it $\widehat{\mathcal{P}}$). Let us
consider now a sequence of natural numbers $\{n: \mbox{\ding{172}}
\}$. It goes from 1 to \ding{172} (note that both numbers, 1 and
\ding{172}, can be expressed by numerals from
$\widehat{\mathcal{P}}$). However, the numeral system
$\widehat{\mathcal{P}}$ is very weak and it allows us    to
observe only ten numbers from the sequence $\{n:
\mbox{\ding{172}}\}$ represented by the following numerals
 \beq \underbrace{1,2}_{finite},
\hspace{5mm} \ldots \hspace{5mm}
\underbrace{\frac{\mbox{\small{\ding{172}}}}{2}-2,
\frac{\mbox{\small{\ding{172}}}}{2}-1,
\frac{\mbox{\small{\ding{172}}}}{2},
\frac{\mbox{\small{\ding{172}}}}{2}+1,
\frac{\mbox{\small{\ding{172}}}}{2}+2}_{infinite}, \hspace{5mm}
\ldots \hspace{5mm} \underbrace{\mbox{\ding{172}}-2,
\mbox{\ding{172}}-1, \mbox{\ding{172}}}_{infinite}.
 \label{4.2.1}
       \eeq
The first two numerals in (\ref{4.2.1}) represent finite numbers,
the remaining eight numerals express infinite numbers, and dots
represent members of the sequence of natural numbers that are not
expressible in~$\widehat{\mathcal{P}}$ and, therefore, cannot be
observed if one uses only this numeral system for this purpose.
\hfill$\Box$
\end{example}

Note that
  Pirah\~{a}     are not able to see  finite numbers larger than 2
using their weak numeral system but these numbers are visible if
one uses a more powerful numeral system. In particular, this means
that when we speak about sets (finite or infinite) it is necessary
to take care about tools used to describe a set. In order to
introduce a set, it is necessary to have a language (e.g., a
numeral system) allowing us to describe both the form of its
elements in a way and the number of its elements. For instance,
the set $A$ from (\ref{4.1.deriva_0}) cannot be defined using the
mathematical language of Pirah\~{a}.

Analogously, the words `the set of all finite numbers' do not
define a set   from our point of view. It is always necessary to
specify which instruments are used to describe (and to observe)
the required set and, as a consequence, to speak about `the set of
all finite numbers expressible in a fixed numeral system'. For
instance, for Pirah\~{a} `the set of all finite numbers'  is the
set $\{1, 2 \}$ and for another Amazonian tribe --
Munduruk\'u\footnote{Munduruk\'u (see \cite{Pica}) fail in exact
arithmetic with numbers larger than   5 but are able to compare
and add large approximate numbers that are far beyond their naming
range. Particularly, they use the words `some, not many' and
`many, really many' to distinguish two types of large numbers (in
this connection think about Cantor's  $\aleph_0$ and $\aleph_1$).}
-- `the set of all finite numbers' is the set $A$ from
(\ref{4.1.deriva_0}). As it happens in Physics, the instrument
used for an observation bounds the possibility of the observation
and determines its accuracy. It is not possible to say what we
shall see during our observation if we have not clarified which
instruments will be used to execute the observation.

Let us consider now infinite sequences as one of the instruments
used by mathematicians to study the   world around us and other
mathematical objects and pro\-cess\-es.  The first   immediate
consequence  of Theorem~\ref{t2}  is that any \textit{sequential}
process can have at maximum \ding{172} elements. This means that a
process of sequential observations    of any object   cannot
contain more than \ding{172} steps\footnote{It is worthy to notice
a deep relation of this observation to the Axiom of Choice. Since
Theorem~\ref{t2} states that any sequence can have at maximum
\ding{172} elements, so this fact holds for the process of a
sequential choice, as well. As a consequence, it is not possible
to choose sequentially more than \ding{172} elements from a set.
This observation also emphasizes the fact that the parallel
computational paradigm is significantly different with respect to
the sequential one because $p$ parallel processes can choose $p
\cdot \mbox{\ding{172}}$ elements from a set.}. Due to
Postulate~1, we are not able to execute any infinite process
physically but we assume the existence of such a process.
Moreover, again due to Postulate~1, only a finite number of
observations of elements of the considered infinite sequence can
be executed by a human who is limited by the numeral system used
for observation. However, the researcher can choose how to
organize the required sequence of observations and which numeral
system to use for it, defining so which elements of the object
he/she can observe. This situation is exactly the same as in
natural sciences: before starting to study a physical object, a
scientist chooses an instrument and its accuracy for the study.

\begin{example}
\label{e4_Turing} Let us consider the set, $\widehat{\mathbb{N}}$,
of extended natural numbers from (\ref{4.2.2}) as an object of our
observation. Suppose that we want to organize the process of the
sequential counting of its elements. Then, due to
Theorem~\ref{t2}, starting from the number 1 this process can
arrive at maximum to \ding{172}. If we consider the complete
counting sequence $\{n: \mbox{\ding{172}}\}$, then we obtain
 \beq
 \begin{array}{l}
              1,2,\hspace{1mm}3,\hspace{1mm}4, \hspace{2mm} \ldots \hspace{2mm}
              \mbox{\ding{172}\tiny$^-$}2,\mbox{\ding{172}\tiny$^-$}1,
              \mbox{\ding{172}}, \mbox{\ding{172}\tiny$^+$}1,
              \mbox{\ding{172}\tiny$^+$}2, \mbox{\ding{172}\tiny$^+$}3, \ldots \\
              \hspace{1mm}\raisebox{3.5ex}{\scalebox{1.0}[1.0]{\rotatebox{180}{$\curvearrowleft$}}}
              \raisebox{3.5ex}{\scalebox{1.0}[1.0]{\rotatebox{180}{$\curvearrowleft$}}}
              \raisebox{3.5ex}{\scalebox{1.1}[1.0]{\rotatebox{180}{$\curvearrowleft$}}}
              \raisebox{3.5ex}{\scalebox{1.1}[1.0]{\rotatebox{180}{$\curvearrowleft$}}}
              \hspace{4.5mm}
              \raisebox{3.5ex}{\scalebox{1.5}[1.0]{\rotatebox{180}{$\curvearrowleft$}}}
              \hspace{1.3mm}
              \raisebox{3.5ex}{\scalebox{1.5}[1.0]{\rotatebox{180}{$\curvearrowleft$}}}
              \hspace{2mm}
              \raisebox{3.5ex}{\scalebox{1.5}[1.0]{\rotatebox{180}{$\curvearrowleft$}}}
              \hspace{0.8mm}
        \vspace*{-4.5mm}\\
              \raisebox{5.5ex}{\scalebox{1}{\rotatebox{0}{$\underbrace{ \hspace{43mm} }_{ \mbox{\ding{172} \small \,steps} } $}}}
 \end{array}
 \label{Turing_3}
       \eeq
In this formula, a more powerful (with respect to
$\widehat{\mathcal{P}}$ from (\ref{4.2.1})) numeral system,
$\widetilde{\mathcal{P}}$, is used. It allows us to see also
numbers three and four through the numerals 3 and 4 and, of
course, such numbers as $\mbox{\ding{172}}-4,
\frac{\mbox{\ding{172}}}{3}, \frac{\mbox{\ding{172}}}{4}-3,$ and
other numbers that can be viewed through numerals obtained as
combinations of symbols   `+', `-', and `/' and numerals 1, 2, 3,
4, and \ding{172} similarly to (\ref{4.2.1}) (we assume that
finite numbers larger than 4 are not expressible in
$\widetilde{\mathcal{P}}$). We omit them in the record
(\ref{Turing_3}) due to a straightforward similarity with
(\ref{4.2.1}).

Analogously, if we start  the process of the sequential counting
from 3, the process   arrives at maximum to $\mbox{\ding{172}}+2$:
\[
\begin{array}{l}
              1,2, 3, \hspace{0.5mm}4, \hspace{2mm} \ldots \hspace{2mm}
              \mbox{\ding{172}\tiny$^-$}2,\mbox{\ding{172}\tiny$^-$}1,
              \mbox{\ding{172}}, \mbox{\ding{172}\tiny$^+$}1,
              \mbox{\ding{172}\tiny$^+$}2, \mbox{\ding{172}\tiny$^+$}3, \ldots \\
              \hspace{8mm}
              \raisebox{3.5ex}{\scalebox{1.1}[1.0]{\rotatebox{180}{$\curvearrowleft$}}}
              \raisebox{3.5ex}{\scalebox{1.1}[1.0]{\rotatebox{180}{$\curvearrowleft$}}}
              \hspace{4.5mm}
              \raisebox{3.5ex}{\scalebox{1.5}[1.0]{\rotatebox{180}{$\curvearrowleft$}}}
              \hspace{1.0mm}
              \raisebox{3.5ex}{\scalebox{1.5}[1.0]{\rotatebox{180}{$\curvearrowleft$}}}
              \hspace{1.5mm}
              \raisebox{3.5ex}{\scalebox{1.5}[1.0]{\rotatebox{180}{$\curvearrowleft$}}}
              \hspace{0.1mm}
              \raisebox{3.5ex}{\scalebox{1.6}[1.0]{\rotatebox{180}{$\curvearrowleft$}}}
              \hspace{0.2mm}
              \raisebox{3.5ex}{\scalebox{1.9}[1.0]{\rotatebox{180}{$\curvearrowleft$}}}
              \vspace*{-4.5mm}\\
              \hspace{7mm}\raisebox{5.5ex}{\scalebox{1}{\rotatebox{0}{$\underbrace{ \hspace{49mm} }_{ \mbox{\ding{172} \small \,steps} } $}}}
 \end{array}
 \]
 The corresponding complete
sequence  used in this case is $\{n+2: \mbox{\ding{172}}\}$. We
can also   change the length of the step in the counting sequence
and consider, for instance, the complete sequence $\{2n-1:
\mbox{\ding{172}}\}$:
\[
 \begin{array}{l}
              1,2,3,4,\hspace{1mm}  \ldots \hspace{1mm}
 \mbox{\ding{172}\tiny$^-$}1,
\mbox{\ding{172}},  \mbox{\ding{172}\tiny$^+$}1,
\mbox{\ding{172}\tiny$^+$}2,
 \hspace{1mm} \ldots \hspace{1mm} 2\mbox{\ding{172}\tiny$^-$}3, 2 \mbox{\ding{172}\tiny$^-$}2, 2\mbox{\ding{172}\tiny$^-$}1, 2 \mbox{\ding{172}},
  2\mbox{\ding{172}\tiny$^+$}1, \ldots \\
              \hspace{1mm}
              \raisebox{3.5ex}{\scalebox{2}[1.1]{\rotatebox{180}{$\curvearrowleft$}}}
              \hspace{-0.5mm}
              \raisebox{3.5ex}{\scalebox{2}[1.1]{\rotatebox{180}{$\curvearrowleft$}}}
               \hspace{2.3mm}
              \raisebox{3.5ex}{\scalebox{2}[1.1]{\rotatebox{180}{$\curvearrowleft$}}}
              \raisebox{3.5ex}{\scalebox{2.8}[1.1]{\rotatebox{180}{$\curvearrowleft$}}}
              \hspace{0.5mm}
              \raisebox{3.5ex}{\scalebox{4}[1.1]{\rotatebox{180}{$\curvearrowleft$}}}
              \hspace{1.0mm}
              \raisebox{3.5ex}{\scalebox{2}[1.1]{\rotatebox{180}{$\curvearrowleft$}}}
              \hspace{0.1mm}
              \raisebox{3.5ex}{\scalebox{5.1}[1.1]{\rotatebox{180}{$\curvearrowleft$}}}
                     \vspace*{-4.5mm}\\
              \raisebox{5.5ex}{\scalebox{1}{\rotatebox{0}{$\underbrace{ \hspace{83mm} }_{ \mbox{\ding{172} \small \,steps} } $}}}
 \end{array}
 \]
If we use again the numeral system  $\widetilde{\mathcal{P}}$,
then among finite numbers it allows us to see only numbers 1 and 3
because already the next number in the sequence, 5, is not
expressible in $\widetilde{\mathcal{P}}$. The last two elements of
the sequence are $2 \mbox{\ding{172}}-3$ and
$2\mbox{\ding{172}}-1$ and $\widetilde{\mathcal{P}}$ allows us to
observe them.  \hfill
 $\Box$
  \end{example}

The introduced definition of a sequence  allows us to work not
only with the first but also with the last element of any sequence
(if they are expressible in the chosen numeral system)
independently whether it has a finite or an infinite number of
elements. Let us use this new definition together with Postulate~2
for studying infinite sets of numerals, in particular, for
calculating the number of points at the interval $[0,1)$ (see
\cite{Sergeyev,informatica}). To do this we need a definition of
the term `point'\index{point} and mathematical tools to indicate a
point. Since this concept is one of the most fundamental, it is
very difficult to find an adequate definition. If we accept (as is
usually done in modern Mathematics) that a \textit{point} $A$
belonging to the interval $[0,1)$ is determined by a numeral $x$,
$x  \in \mathbb{S},$ called \textit{coordinate of the point A}
where $\mathbb{S}$ is a set of numerals,    then we can indicate
the point $A$ by its coordinate  $x$  and we are able to execute
the required calculations.

It is worthwhile to emphasize  that we have not postulated that
$x$ belongs to the   set, $\mathbb{R}$, of real numbers  as it is
usually done, because we can express coordinates only by numerals
and different choices of numeral systems lead to various sets of
numerals. This situation   is a direct consequence of Postulate~2
and is typical for natural sciences where it is well known that
instruments influence the results of observations. Remind again
the work with a microscope: we decide the level of the precision
we need and obtain a result which is dependent on the chosen level
of accuracy. If we need a more precise or a more rough answer, we
change the lens of our microscope.

We should decide now which numerals we shall use to express
coordinates of the points. After this choice we can calculate the
number of numerals expressible in this system and, as a result, we
obtain the number of points at the interval $[0,1)$. Different
variants (see \cite{Sergeyev,informatica}) can be chosen depending
on the precision level we want to obtain.  For instance,   we can
choose a positional numeral system with a finite radix $b$ that
allows us to work with numerals
 \beq
(0.a_{-1} a_{-2}  \ldots a_{-(\mbox{\tiny\ding{172}}-1)}
a_{-\mbox{\tiny\ding{172}}})_b, \hspace{5mm}  a_{-i} \in \{ 0, 1,
\ldots b-2, b-1 \}, \hspace{3mm}  1 \le i \le \mbox{\ding{172}}.
 \label{3.103calcolo}
       \eeq
Then,   the number of numerals (\ref{3.103calcolo}) gives us the
number of points within the interval $[0,1)$ expressed by these
numerals. Note that  a number using  the positional numeral system
(\ref{3.103calcolo}) cannot have more than grossone digits
(contrarily to sets discussed in Example~\ref{e4_Turing}) because
a numeral having $g>\mbox{\ding{172}}$ digits would not be
observable in a sequence. In this case such a record  becomes
useless in
 sequential computations because it does not allow one
to identify  numbers since $g-\mbox{\ding{172}}$ numerals remain
non observed.

\begin{theorem}
\label{t1_Turing} If  coordinates of   points $x \in [0,1)$ are
expressed by numerals (\ref{3.103calcolo}), then the number of the
points $x$ over $[0,1)$  is equal to $b^{\mbox{\tiny\ding{172}}}$.
\end{theorem}

\textit{Proof.} In the numerals (\ref{3.103calcolo})   there is a
sequence  of digits,
 $a_{-1} a_{-2} \ldots
a_{-(\mbox{\tiny\ding{172}}-1)} a_{-\mbox{\tiny\ding{172}}}$, used
to express   the fractional part of the number. Due to the
definition of the sequence and Theorem~\ref{t2}, any infinite
sequence can have at maximum \ding{172} elements. As a result,
there is \ding{172} positions   on the right of the dot that can
be filled in by one of the $b$ digits from the
alphabet\index{alphabet} $\{ 0, 1, \ldots , b-1 \}$. Thus, we have
$b^{\mbox{\tiny\ding{172}}}$ combinations to express the
fractional part of the number. Hence,   the positional numeral
system using the numerals of the form (\ref{3.103calcolo}) can
express $b^{\mbox{\tiny\ding{172}}}$ numbers. \hfill
 $\Box$
\begin{corollary}
\label{c1_Turing}

 The  number of   numerals
  \beq
(a_{1} a_{2} a_{3}  \ldots
a_{\mbox{\tiny\ding{172}}-2}a_{\mbox{\tiny\ding{172}}-1}
a_{\mbox{\tiny\ding{172}}})_b, \hspace{5mm} a_i \in \{ 0, 1,
\ldots b-2, b-1 \}, \hspace{5mm}  1 \le i \le \mbox{\ding{172}},
 \label{Turing_4}
       \eeq
expressing integer numbers in the positional system with a finite
radix $b$ in the alphabet $\{ 0, 1, \ldots b-2, b-1 \}$ is equal
to $b^{\mbox{\tiny\ding{172}}}$.
\end{corollary}

\textit{Proof.}  The  proof is a straightforward consequence of
Theorem~\ref{t1_Turing} and is so omitted. \hfill $\Box$

\begin{corollary}
\label{c2_Turing}

  If  coordinates of   points $x \in (0,1)$ are
expressed by numerals (\ref{3.103calcolo}), then the number of the
points $x$ over $(0,1)$  is equal to
$b^{\mbox{\tiny\ding{172}}}-1$.
\end{corollary}

\textit{Proof.}  The  proof follows immediately from
Theorem~\ref{t1_Turing}. \hfill
 $\Box$

Note that Corollary \ref{c2_Turing} shows that  it becomes
possible now to observe and to register the difference of the
number of elements of two infinite sets (the interval $[0,1)$ and
the interval $(0,1)$, respectively) even when only one element
(the point 0) has been excluded from the first set in order to
obtain the second one.

\section{The Turing machines viewed through the lens\\
of the new  methodology}
\label{s3_Turing}

In the previous section, we studied static infinite mathematical
objects -- sets -- by using infinite sequences as tools of the
research. Let us establish now what can we say with respect to
physical and mathematical processes viewed as objects of
observation having in mind the triad `object, instrument, and
researcher' emphasized by Postulate~2. Our main attention will be
focused on processes related to the Turing machines and various
manifestations of infinity taking place during the  work of the
machines and during mathematical descriptions of  the machines
performed by researchers.

Remind that traditionally, a Turing machine (see, e.g.,
\cite{Hopcroft_Ullman,Turing}) can be defined as a 7-tuple
 \beq
\M=\seq{Q, \Gamma, \bar{b}, \Sigma, q_0, F, \delta},
\label{Turing_7}
 \eeq
 where $Q$ is
a finite and not empty set of states; $\Gamma$ is a finite set of
symbols; $\bar{b}\in\Gamma$ is a symbol called blank; $\Sigma
\subseteq \{\Gamma-{\bar{b}}\}$ is the set of input/output
symbols; $q_0\in Q$ is the initial state; $F\subseteq Q$ is the
set of final states; $\delta: \{Q-F\}\times\Gamma\mapsto
Q\times\Gamma\times\{R,L,N\}$ is a partial function called the
transition function, where $L$ means left, $R$ means right, and
$N$ means no move.

Specifically, the machine is supplied with: (i) a \textit{tape}
running through it which is divided into cells each capable of
containing a symbol $\gamma\in\Gamma $, where $\Gamma$ is called
the tape alphabet, and $\bar{b}\in\Gamma$ is the only symbol
allowed to occur on the tape infinitely often; (ii) an
\textit{head} that can read and write symbols on the tape and move
the tape left and right one  and only one  cell at a time. The
behavior of the machine  is specified by its \textit{transition
function} $\delta$ and consists of a sequence of  computational
steps; in each step the machine reads the symbol under the head
and applies the \textit{transition function} that, given the
current state of the machine and the symbol it is reading on the
tape, specifies (if it is defined for these inputs): (i) the
symbol $\gamma\in\Gamma$ to write on the cell of the tape under
the head; (ii) the move of the tape ($L$ for one cell left, $R$
for one cell right, $N$ for no move); (iii) the next state $q\in
Q$ of the machine.

Following Turing (see \cite{Turing}), we consider  machines that
have finite input and output alphabets, inputs of a finite length,
a finite number of internal states but can work an infinite time
and are able to produce outputs of an infinite length. Hereinafter
such a machine is called an imaginary Turing machine,
$\mathcal{T^I}$. In order to study the limitations of practical
automatic computations, we also consider machines,
$\mathcal{T^P}$, that can be constructed physically. They are
identical to $\mathcal{T^I}$ but are able to work only a finite
time and can produce only finite outputs.  We study both kinds of
machines:
\begin{itemize}
  \item[-] from the point of view of
their outputs called by Turing `computable numbers' or `computable
sequences';
  \item[-] from the point of view of algorithms that can be
  executed by a Turing machine.
\end{itemize}

\subsection{Computable
sequences} \label{s3.1_Turing}

Let us consider first a physical machine $\mathcal{T^P}$. We
suppose that its output is written on the tape using an alphabet
$\Sigma$ containing symbols $\{ 0, 1, \ldots b-2, b-1 \}$ where
$b$ is a finite number (Turing in \cite{Turing} uses $b=10$).
Thus, the output consists of a sequence of digits that can be
viewed as a number in a positional system $\mathcal{B}$ with the
radix $b$.

It follows from Postulate~1 (reflecting a fundamental law existing
in the real world) that $\mathcal{T^P}$ should stop after a finite
number of iterations. The magnitude of this value depends on the
physical construction of the machine, the way the notion
`iteration' has been defined, etc., but in any case this number is
finite. The machine stops in two cases: (i) it has finished
execution of its program and stops; (ii) it has not finished
execution of the program and stops just because of a breakage of
some of its components. In both cases the output sequence
 \beq
  (a_{1} a_{2}
a_{3}  \ldots a_{k-1},a_{k})_b,  \hspace{5mm} a_i \in \{ 0, 1,
\ldots b-2, b-1 \},  \hspace{3mm}  1 \le i \le k,
 \label{Turing_5}
       \eeq
 of $\mathcal{T^P}$
has a finite length $k$. Suppose that the maximal length of the
output sequence that can be computed by $\mathcal{T^P}$ is equal
to a finite number $K_{\mathcal{T^P}}$. Then it follows $k \le
K_{\mathcal{T^P}}$. This means that there exist
 problems that cannot be solved by $\mathcal{T^P}$
if the length of the output  necessary to write down the solution
outnumbers $K_{\mathcal{T^P}}$. If a machine $\mathcal{T^P}$ has
stopped to write the output after it has printed
$K_{\mathcal{T^P}}$  symbols then it is not clear whether the
obtained output is a solution or just a result of the depletion of
its computational resources. In particular, with respect to the
halting problem it follows that all algorithms stop on
$\mathcal{T^P}$.

Let us call a person working with the machine and reading the
output as a \textit{researcher} (or a  \textit{user}). Then, in
order to be able to read and to understand the output, the
researcher should have his/her own positional numeral system
$\mathcal{U}$ with an alphabet $\{ 0, 1, \ldots u-2, u-1 \}$ where
$u \ge b$ from (\ref{Turing_5}). Otherwise, the output cannot be
understood and decoded by the user. Moreover, he/she should be
able to read and to interpret output sequences of symbols with the
length $K_{\mathcal{U}} \ge K_{\mathcal{T^P}}$. If the situation
$K_{\mathcal{U}} < K_{\mathcal{T^P}}$ holds, then this means that
the user is not able to interpret the obtained result. Thus, the
number $K^*=\min \{K_{\mathcal{U}}, K_{\mathcal{T^P}}\}$ defines
the length of the outputs that can be computed and then observed
and interpreted by the user. As a consequence, algorithms
producing outputs having more than $K^*$ positions become less
interesting from the practical point of view.

It is possible to make analogous considerations with respect to
alphabets and numeral systems used for input sequences restricting
so again the number of algorithms useful from the practical point
of view. Finally, the algorithm should be written down someway.
This operation is executed by using an alphabet and a numeral
system used for writing down the algorithm     introduces
limitations to the algorithms that can be proposed for executing
them on a machine. These considerations are important because on
the one hand, they establish limits of practical  automatic
computations and on the other hand, they emphasize the role of
numeral systems in codifying   algorithms and interpreting results
of computations.

Let us turn now to   imaginary Turing machines $\mathcal{T^I}$.
Such a machine can produce outputs (\ref{Turing_5}) with an
infinite number of symbols $k$. In order to be \textit{observable
in a sequence}, an output should have $k \le \mbox{\ding{172}}$
(remind that the positional numeral system $\mathcal{B}$ includes
numerals being a \textit{sequence} of digits, in order to be a
numeral, the output should have $k \le \mbox{\ding{172}}$).
Outputs observable in a sequence play an important role in the
further consideration.

\begin{theorem}
\label{t2_Turing} Let $M$ be the number of all possible complete
computable sequences  that can be produced by    imaginary Turing
machines using outputs (\ref{Turing_5}) being numerals in the
positional numeral system $\mathcal{B}$. Then it follows $M  \le
b^{\mbox{\tiny\ding{172}}}$.
\end{theorem}

\textit{Proof.}  This result follows from   the definitions of the
complete sequence and the positional numeral system considered
together with Theorem~\ref{t1_Turing}  and
Corollary~\ref{c1_Turing}. \hfill
 $\Box$
 \begin{corollary}
\label{c3_Turing} Let us consider an imaginary Turing machine
$\mathcal{T^{I}}$ working with the alphabet $\{0,1,2\}$ and
computing the following complete computable sequence
 \beq
\underbrace{0,1,2, 0,1,2, 0,1,2,     \hspace{1mm} \ldots
\hspace{1mm} 0,1,2, 0,1,2 }_{\mbox{\ding{172} positions}}.
 \label{Turing_6}
       \eeq
Then   imaginary Turing machines working with the output alphabet
$\{0,1\}$ cannot   produce observable in a sequence outputs that
codify and compute (\ref{Turing_6}).
\end{corollary}

\textit{Proof.} Since the numeral 2 does not belong to the
alphabet $\{0,1\}$ it should be coded by more than one symbol. One
of the coding using the minimal number of symbols in the alphabet
$\{0,1\}$ necessary to code numbers $0,1,2$ is $\{00,01,10\}$.
Then the output corresponding to (\ref{Turing_6}) and computed in
this codification should be
 \beq
  00,01,10,
00,01,10, 00,01,10,     \hspace{1mm} \ldots \hspace{1mm} 00,01,10,
00,01,10.
 \label{Turing_6.1}
       \eeq
Since the  output (\ref{Turing_6}) contains grossone positions,
the  output (\ref{Turing_6.1}) would contain $2\mbox{\ding{172}}$
positions.  However, in order to be observable in a sequence,
(\ref{Turing_6}) should not have more than grossone positions.
This fact completes the proof.
 \hfill
 $\Box$

At first glance results established by Theorem~\ref{t2_Turing} and
Corollary \ref{c3_Turing}    sound quite unusual  for a person who
studied the behavior of Turing machines on infinite computable
sequences using traditional mathematical tools. However, they do
not contradict   each other. Theorem~\ref{t2_Turing} and Corollary
\ref{c3_Turing} do not speak about \textit{all} Turing machines.
They consider only those machines that produce complete output
sequences. If the object of observation (in this case -- the
output) contains more than grossone elements, it cannot be
observed and, therefore, is less interesting from the point of
view of practical computations.

It is important to emphasize that these results are in line with
the situation that we have in the real world with a finite number
of positions in the output sequences. For instance, suppose that a
physical Turing machine $\mathcal{T^P}$  has 6 positions at its
output, the numeral system $\{0,1,2\}$, and the sequence $0,1,2,
0,1,2$ is computed. Then there does not exist a Turing machine
working with the output alphabet  $\{0,1\}$ able to calculate the
sequence $0,1,2, 0,1,2$ using the output having 6 positions.

In order to understand Theorem~\ref{t2_Turing} and
Corollary~\ref{c3_Turing} better, let us return to the Turing
machine as it has been described in~\cite{Turing} and comment upon
connections between the traditional results  and the new ones.
First, it is necessary to mention that results of Turing and
results of Theorems~\ref{t1_Turing},~\ref{t2_Turing}, and
Corollary~\ref{c3_Turing}  have been formulated using different
mathematical languages. The one used by Turing has been developed
by Cantor and did not allow Turing to distinguish within continuum
various sets having  different number of elements. The new numeral
system using grossone allows us to do this.

Cantor has proved, by using his famous diagonal argument, that the
number of elements of the set $\mathbb{N}$ is less than the number
of real numbers at the interval $[0,1)$ \textit{without
calculating the latter}. To do this, he expressed  real numbers in
a positional numeral system. We have shown that this number will
be different depending on the radix $b$ used in the positional
system to express real numbers. However, all of the obtained
numbers, $b^{\mbox{\tiny\ding{172}}}$, are larger than the number
of elements of the set of natural numbers, \ding{172}.

Thus, results presented in Theorem~\ref{t2_Turing} and
Corollary~\ref{c3_Turing} should be considered just as a more
precise analysis of the situation related to the existence of
different infinities discovered by Cantor. The usage of a more
powerful numeral system gives a possibility to distinguish and to
describe more mathematical objects within the continuum, in the
same way as the usage of a stronger lens in a microscope gives a
possibility to distinguish more objects within an   object that
seems to be indivisible when viewed by a weaker lens.

As a consequence,  the mathematical results obtained by Turing and
those presented in Theorems~\ref{t1_Turing},~\ref{t2_Turing}, and
Corollary~\ref{c3_Turing}  do not contradict   each other.
\textit{They   are correct with respect to mathematical languages
used to express them and correspond to different accuracies of the
observation.} Both mathematical languages observe and describe the
same object -- computable sequences -- but with different
accuracies. This fact is one of the manifestations of the
relativity of mathematical results formulated by using different
mathematical languages.

Another manifestation of this relativity is obviously related to
the concept of the universal Turing machine and to the process of
establishing equivalence between machines. Notice that
Theorem~\ref{t2_Turing} and Corollary~\ref{c3_Turing} emphasize
dependence of the outputs of Turing machines on a finite alphabet
$\{ 0, 1, \ldots b-2, b-1 \}$ used for writing down computable
sequences. Therefore, when a researcher describes a Turing
machine, there exists the dependence  of the description on the
finite numeral system used by the researcher. First, the
description  is limited by   alphabets $\{ 0, 1, \ldots b-2, b-1
\}$ known to the humanity at the present situation. Second, by the
maximal length, $K_{\mathcal{U}}$, of the sequence of symbols
written in the fixed alphabet that the researcher is able to read,
to write, and to understand.

It is not possible to describe a Turing machine (the object of the
study) without the usage of a numeral system (the instrument of
the study). Our possibilities   to observe and to describe  Turing
machines and to count their number are limited by the numeral
systems known to the humanity at the moment. Again, as it happens
in natural sciences, the tools used in the study limit the
researcher. As a result, it becomes not possible to speak about an
absolute number of \textit{all possible Turing machines
$\mathcal{T^I}$}. It is always necessary to speak about the number
of all possible Turing machines $\mathcal{T^I}$ expressible in a
fixed numeral system (or in a group of them).

The same limitations play   an important role in the process of
simulating one machine $\mathcal{T^I}$ by another.  In order to be
able to execute this operation it is necessary to calculate the
respective description number (see \cite{Turing}) and this will be
possible only for description numbers expressible in the  finite
alphabets known at the current moment to the researcher and the
length of these numbers will be limited by the number
$K_{\mathcal{U}}$. A machine $\mathcal{T^I}$ having the
description number not satisfying these constraints cannot be
simulated because the instrument -- a numeral system -- required
for such re-writing is not powerful enough (as usual, devil is in
the details).

Let us consider now  from positions of the new numeral system
including gross\-one the situation related to the  enumerability
of machines $\mathcal{T^I}$   studied by Turing.

\begin{theorem}
\label{t3_Turing} The maximal number of complete computable
sequences produced by   imaginary Turing machines  that can be
enumerated in a sequence   is equal to~$\mbox{\ding{172}}$.
\end{theorem}

\textit{Proof.}  This result follows from the definition of a
complete sequence. \hfill
 $\Box$

Let us consider the results of Theorems~\ref{t2_Turing}
and~\ref{t3_Turing} together. Theorem~\ref{t2_Turing} gives an
upper bound for the number of complete computable sequences that
can be computed using a fixed radix $b$. However, we do not know
how many of $b^{\mbox{\tiny\ding{172}}}$ sequences can be results
of computations of a Turing machine. Turing establishes that their
number is enumerable. In order to obtain this result, he used the
mathematical language developed by Cantor and this language did
not allow him to distinguish sets having different infinite
numbers of elements, e.g., in the traditional language that he
used the sets of even, natural, and integer numbers all are
enumerable.

The introduction of grossone gives a possibility to execute a more
precise analysis and to determine that these sets have different
numbers of elements: $\frac{\mbox{\ding{172}}}{2},
\mbox{\ding{172}},$ and $2\mbox{\ding{172}}+1$, respectively. If
the number of complete computable sequences, $M_{\mathcal{T^I}}$,
is larger than grossone, then there can be different sequential
enumerating processes  that enumerate  complete computable
sequences in different ways. Theorem~\ref{t3_Turing} states that,
in any case, each of these enumerating sequential processes cannot
contain more than grossone members.

We conclude this subsection by  noticing  that the results
presented in it establish limitations for the number of computable
sequences  only from the point of view of the output sequences and
their alphabets. An analogous analysis can be done  using
limitations imposed by the number of states of Turing machines,
their inputs, and the respective finite alphabets, as well.

\subsection{Processes of automatic computations and their
descriptions} \label{s3.2_Turing}

First, we take  notice that if we want to observe a process of
computations~$A$ performed by a Turing machine ($\mathcal{T^P}$ or
$\mathcal{T^I}$) while it executes an algorithm, then we   do it
by executing observations of the machine in a sequence of moments.
In fact, it is not possible to organize a \textit{continuous}
observation of the machine. Any instrument used for an observation
has its accuracy and there will always be a minimal period of time
related to this instrument allowing one to distinguish two
different moments of time and, as a consequence, to observe (and
to register) the states of the object in these two moments. In the
period of time passing between these two moments the object
remains unobservable.

Hence, the observations are made in a sequence (that is  an
instrument of the research) and the process of computations $A$ is
the object of the study. In the simplest case we observe $A$ only
two times: at the starting point when we supply the input data and
at the ending point of the process of computation when we read the
results.  In alternative,  observations are made to look at
intermediate results or even at particular moves of the parts of
the machine (e.g., reading   a symbol, writing  a symbol, etc.).

On the one hand, since our observations are made in a sequence, it
follows from Theorem~\ref{t2} that
 the process of observations can have at maximum
\ding{172} elements. This means that inside a computational
process it is possible to fix  more than grossone steps (defined
someway) but it is not possible to count them one by one in a
sequence containing more than grossone elements. For instance, in
a time interval $[0,1)$, numerals (\ref{3.103calcolo}) can be used
to identify moments of time but not more than grossone of them can
be observed in a sequence.

On the other hand, it is important to stress  that any process
itself, considered independently on the researcher, is not
subdivided in iterations, intermediate results, moments of
observations, etc. This is a direct consequence of Postulate~2,
the consequence that is also in line with  the Sapir--Whorf
thesis, particularly, with results of Whorf (see~\cite{Whorf})
related to his analysis of the differences between Western
languages  and the Hopi language (a Uto-Aztecan language spoken by
the Hopi people of northeastern Arizona, USA). Analyzing the
relationship between language, thought, and reality in these two
types of languages  (see also recent experimental data  and the
relative discussion in \cite{Gumperz_Levinson,Lucy}) Whorf raises
a barrier between them.

Western languages  tend to analyze reality as objects in space.
There exist other languages, including many Native American
languages, that are oriented towards processes. To monolingual
speakers of such languages, the  constructions of Western
languages related to objects and separate events may make little
sense. On the other hand, due to Whorf, the relativistic physics
-- a subject being very hard for understanding for a Western
language speaker --  a Hopi speaker would find fundamentally
easier to grasp. Whorf writes:
\begin{quote}
 We dissect nature along lines laid down by our native language.
The categories and types that we isolate from the world of
phenomena we do not find there because they stare every observer
in the face; on the contrary, the world is presented in a
kaleidoscope flux of impressions which has to be organized by our
minds -- and this means largely by the linguistic systems of our
minds. We cut nature up, organize it into concepts, and ascribe
significances as we do, largely because we are parties to an
agreement to organize it in this way -- an agreement that holds
throughout our speech community and is codified in the patterns of
our language $[\ldots]$
\end{quote}

The Sapir--Whorf  thesis is interesting for us because, in a
complete  accordance with our methodological  positions, it
separates the object of observations from its representation by
one or another language.  In particular, with respect to automatic
computations we emphasize that a machine (a physical or an
imaginary one) executing a computation does not distinguish an
importance of one moment during the execution of an action with
respect to another and does not count them. Certain milestones
inside a process (computational steps defined someway, operations,
iterations, etc.) are introduced from outside of the studied
process by the researcher because these specific points are
interesting for the observer for some reasons and can be expressed
in his language. The notion `sequence' is a tool invented by human
beings, it is a part of the modern mathematical languages
(developed mainly in the frame of Western languages dissecting
processes in separate events), it does not take part of the object
of the study.

When we speak about a computer executing iterations of a certain
algorithm, \textit{we} subdivide the process of computations on
iterations and \textit{we}   count them. As a result, it is
necessary to speak   about the computational  power of computers
(in particular, of Turing machines) coupled with our possibilities
to use them, to follow computational processes, to be able to
provide input data, and to read results of computations. The
understanding of the fact that computations executed by a computer
and our observations and descriptions of these computations are
different processes lead  to the necessity to rethink such notions
as \textit{iteration}  and \textit{algorithm}.

At the moment when we decide what is an iteration of our
algorithm, we are choosing the instrument of our investigation and
the further results will depend on the chosen accuracy (or
granularity) of observations. For instance, with respect to Turing
machines an iteration can be a single operation of the machine
such as reading a symbol from the tape, or moving the tape, etc.
Another possible example of such a choice (that is usually used in
the Computer Science literature) is to observe the machine when
its configuration has been changed. All these choices produce
different sequences of observations that form an algorithm if we
add to them an input and an output being the symbols present on
the tape of $\mathcal{T^I}$ at the first and at the last
observation, respectively.

In order to conclude our discussion on the notion of the algorithm
it is necessary to remind that   any sequence cannot contain more
then grossone steps. Thus, after we have chosen what is the
iteration of our algorithm, the maximal number of these iterations
cannot outnumber \ding{172}. As usual, the choice of the numeral
system used to describe  iterations and their results determines
what will be observable for the researcher. Similarly, the choice
of the numeral system (and, in general, of the mathematical
language) used to describe the algorithm will limit the type of
the algorithms that can be described.

The notion of the result of a computation on $\mathcal{T^I}$ has
the same sense as it was for $\mathcal{T^P}$.   If $\mathcal{T^I}$
has not stopped after   \ding{172} observations, then this means
that we have finished our possibilities of observations and we
cannot say whether the symbols present at the tape during this
observation are effectively the solution to the problem.   By a
complete analogy with  $\mathcal{T^P}$, computations finish either
because the machine $\mathcal{T^I}$ stops or because we are not
any more able to observe computations (since $\mathcal{T^I}$ is an
imaginary one, the possibility of its breakage is not taken into
consideration). In particular, this means that with respect to the
halting problem all algorithms stop but this does not mean that
the obtained result is a solution.

The analysis given above shows us that it is not possible to speak
about the computational power of a Turing machine without taking
into consideration a number of limitations   introduced by the
languages. Among them there are at least the following: the
language used to describe the algorithm and iterations it consists
of; the language used to describe the process of computations; and
the language used to describe the Turing machine itself (its input
and output alphabets, its states, etc.). Notice that this
situation with a description of automatic computations is just a
particular case of the situation emphasized by Postulate~2: when
we study an object it is necessary to be aware of the accuracy and
the capability of instruments used for the study.

The obtained picture of the computability is significantly richer
and complex with respect to traditional views (see
\cite{Church,Barry_Cooper,Davis,Kleene,Kolmogorov,Kolmogorov_Uspensky,Markov,Post,Turing}).
The  classic  Turing theory contains   a number of theoretical
results showing the same computational power of different variants
of Turing machines and establishing that the differences among
machines  $\mathcal{T_{\mbox{\tiny 1}}^{I}}$ and
$\mathcal{T_{\mbox{\tiny 2}}^{I}}$ result only in the different
number of steps that will be necessary to each machine for
computing the required output. Some of the limitations on this
point of view have been already discussed in
Section~\ref{s3.1_Turing}. In this section, we have emphasized a
number of additional limitations. Again, as it was in
Section~\ref{s3.1_Turing}, the difference with the traditional
results is not a contradiction. These differences arise  because
the mathematical language used for these traditional studies did
not allow people to see the differences among various models of
computations.

\section{Usage of traditional and new languages for comparing deterministic and non-deterministic Turing machines}
 \label{s4_Turing}

In order to illustrate the new way of reasoning, let us discuss
the traditional and   new results regarding the computational
power of deterministic and non-deterministic Turing machines. For
simplicity, we do not   take into consideration limitations
described in Section~\ref{s3.1_Turing}. Let us first remind the
traditional point of view.

A non-deterministic Turing machine (see~\cite{Hopcroft_Ullman})
can be   defined (cf. (\ref{Turing_7}))   as a 7-tuple
 \beq
  \M_{N}=\seq{Q, \Gamma, \bar{b},
\Sigma, q_0, F, \delta_{N}},
 \label{Turing_8}
 \eeq
 where $Q$ is a finite and not empty
set of states; $\Gamma$ is a finite set of symbols;
$\bar{b}\in\Gamma$ is a symbol called blank; $\Sigma \subseteq
\{\Gamma-{\bar{b}}\}$ is the set of input/output symbols; $q_0\in
Q$ is the initial state; $F\subseteq Q$ is the set of final
states; $\delta_{N}: \{Q-F\}\times\Gamma\mapsto \mathcal{P}
(Q\times\Gamma\times\{R,L,N\})$ is a partial function called the
transition function, where $L$ means left, $R$ means right, and
$N$ means no move.

As for a  deterministic  Turing machine (see (\ref{Turing_7})),
the behavior of $\M_{N}$ is specified by its  transition function
$\delta_{N}$ and consists of a sequence of computational steps. In
each step, given the current state of the machine and the symbol
it is reading on the tape, the  transition function  $\delta_{N}$
returns (if it is defined for these inputs) a set of triplets each
of which specifies: (i) a symbol $\gamma\in\Gamma$ to write on the
cell of the tape under the head; (ii) the move of the tape ($L$
for one cell left, $R$ for one cell right, $N$ for no move); (iii)
the next state $q\in Q$ of the Machine. Thus, in each
computational step, the machine can \textit{non-deterministically}
execute different computations, one for each triple returned by
the transition function.

An important characteristic of a non-deterministic Turing machine
(see, e.g.,~\cite{Ausiello}) is its non-deterministic degree
\[
d=\nu(\M_{N})=\max_{q\in Q-F,\gamma \in \Gamma} \abs{\delta_N
(q,\gamma)}
  \]
  defined as the maximal number of
different configurations reachable in a single computational step
starting from a given configuration. The behavior of the machine
can be then represented as a tree whose branches are the
computations that the machine can execute starting from the
initial configuration represented by the node 0  and nodes of the
tree at the levels 1, 2, etc. represent  subsequent configurations
of the machine.

 \begin{figure}[t]
  \begin{center}
    \epsfig{ figure = 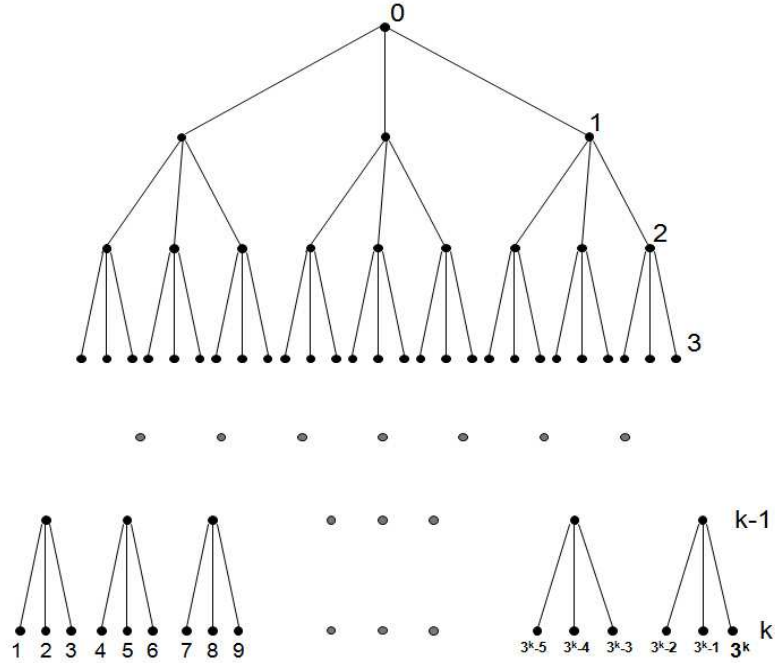, width = 4in, height = 3.5in,  silent = yes }
    \caption{The computational tree of a non-deterministic Turing machine
$\M_{N}$ having the non-deterministic degree $d = 3$}
 \label{Figure_Turing}
  \end{center}
\end{figure}

 Let us consider  an example  shown in
Fig.~\ref{Figure_Turing} where a non-deterministic machine
$\M_{N}$ having the non-deterministic degree $d = 3$  is
presented. The depth of the computational tree is equal to $k$. In
this example,   it is supposed that the computational tree of
$\M_{N}$  is complete (i.e., each node has exactly $d$ children).
Then, obviously, the computational tree of $\M_{N}$ has $d^{k} =
3^k$ leaf nodes.

An important result for the classic theory on Turing machines (see
e.g.,  \cite{Ausiello}) is that for any non-deterministic Turing
machine $\M_{N}$ there exists an equivalent deterministic Turing
machine $\M_{D}$. Moreover, if the depth of the computational tree
generated by $\M_{N}$ is equal to $k$,  then for simulating
$\M_{N}$, the deterministic machine $\M_{D}$ will execute at most
 \[
 K_{\M_{D}}=\sum_{j=0}^k jd^j=O(kd^k)
 \]
 computational steps.

Intuitively, for simulating $\M_{N}$, the deterministic Turing
machine $\M_{D}$ executes a breadth-first visit of the
computational tree of $\M_{N}$. If we consider the  example  from
Fig.~\ref{Figure_Turing} with $k=3$, then the computational tree
of $\M_{N}$ has $d^{k} = 27$ leaf nodes and $d^{k} = 27$
computational paths consisting of $k=3$ branches (i.e.,
computational steps). Then, the tree contains
 $d^{k-1} = 9$ computational paths
consisting of $k-1=2$ branches and $d^{k-2} = 3$ computational
paths consisting of $k-2=1$ branches. Thus, for simulating all the
possible computations of $\M_{N}$, i.e., for   complete visiting
the computational tree of $\M_{N}$ and considering all the
possible computational paths of $j$ computational steps for each
$0\leqslant{j}\leqslant k$, the deterministic Turing machine
$\M_{D}$ will execute $K_{\M_{D}}$ steps. In particular,  if
$\M_{N}$ reaches a final configuration (e.g., it accepts a string)
in $k\geqslant0$ steps and if $\M_{D}$ could consider only the
$d^{k}$ computational paths which consist of $k$ computational
steps, it will executes at most $kd^{k}$ steps for reaching this
configuration.

These results show an exponential growth of the time required for
reaching a final configuration by the deterministic Turing machine
$\M_{D}$ with respect to the time required by the
non-deterministic Turing machine $\M_{N}$, assuming that the time
required for both machines for a single step is the same. However,
in the classic theory on Turing machines it is not known if there
is a more efficient simulation of $\M_{N}$. In other words, it is
an important and open problem of Computer Science theory to
demonstrate that it is not possible to simulate a
non-deterministic Turing machine by a deterministic Turing machine
with a sub-exponential numbers of steps.

Let us now return to the new mathematical language. Since the main
interest to machines (\ref{Turing_8}) is related to their
theoretical properties, hereinafter we start by a comparison of
imaginary deterministic Turing machines, $\mathcal{T^I}$, with
imaginary machines $\M_{N}$ from (\ref{Turing_8}) denoted as
$\mathcal{T^{IN}}$. Physical machines $\mathcal{T^P}$ and
$\mathcal{T^{PN}}$  are considered at the end of this section.

Due to the analysis made in Section~\ref{s3.2_Turing}, we should
choose the accuracy (granularity) of  processes of observation of
both machines, $\mathcal{T^{I}}$ and $\mathcal{T^{IN}}$. In order
to be close as much as possible to the traditional results, we
consider an application of the transition function of the machine
as our observation granularity. With respect to $\mathcal{T^{IN}}$
this means that the nodes of the computational tree are observed.
With respect to $\mathcal{T^{I}}$ we consider sequences of such
nodes. For both cases the initial configuration is not observed,
i.e., we start our observations from level 1 of the computational
tree.

This choice of the observation granularity is particularly
attractive due to its accordance with the   traditional
definitions of Turing machines (see definitions (\ref{Turing_7})
and (\ref{Turing_8})). A more fine granularity of  observations
allowing us to follow internal operations of the machines  can be
also chosen but is not so convenient. In fact, such an accuracy
would mix internal operations of the machines with operations of
the algorithm that is executed. A   coarser granularity could be
considered, as well. For instance, we could define as a
computational step two consecutive applications of the transition
function of the machine. However, in this case we do not observe
all the nodes of the computational tree. As a consequence, we
could miss some results of the computation as the machine could
reach a final configuration before completing an observed
computational step and we are not able to observe when and on
which configuration the machine stopped. Then, fixed the chosen
level of granularity the following result holds immediately.

\begin{theorem}
\label{t4_Turing}   (i) With the chosen level of granularity no
more than \ding{172} computational steps of the machine
$\mathcal{T^{I}}$ can be observed in a sequence. (ii) In order to
give possibility to observe at least one computational path  of
the computational tree of $\mathcal{T^{IN}}$ from the level ~1 to
the level~$k$, the depth, $k \ge 1$, of the computational tree
cannot be larger than grossone, i.e., $k \le \mbox{\ding{172}}$.
\end{theorem}

\textit{Proof.}  Both results follow from the analysis made in
Section~\ref{s3.2_Turing} and Theorem~\ref{t2}. \hfill
 $\Box$

\begin{corollary}
\label{c4_Turing} Suppose that $d$ is the non-deterministic degree
of $\mathcal{T^{IN}}$ and $S$ is the number of leaf nodes of the
computational tree with a depth $k$ representing the possible
results of the computation of $\mathcal{T^{IN}}$. Then it is not
possible to observe all $S$ possible results of the computation of
$\mathcal{T^{IN}}$ if the computational tree  of
$\mathcal{T^{IN}}$ is complete and $d^{k} >$\ding{172}.
\end{corollary}

\textit{Proof.} For the number of leaf nodes of the tree, $S$, of
a generic non-deterministic Turing machine $\mathcal{T^{IN}}$ the
estimate $S \le d^{k}$ holds. In particular, $S=d^{k}$ if the
computational tree is complete, that is our case. On the other
hand, it follows from Theorem~\ref{t2} that any sequence of
observations cannot have more than grossone elements. As a
consequence, the same limitation holds for the sequence of
observations of the leaf nodes of the computational tree. This
means that we are not able to observe all the possible results of
the computation of our non-deterministic Turing machine
$\mathcal{T^{IN}}$ if $d^{k} >$\ding{172}.
   \hfill  $\Box$

\begin{corollary}
\label{c5_Turing} Any sequence of observations of the nodes of the
computational tree of a non-deterministic Turing machine
$\mathcal{T^{IN}}$  cannot observe all the nodes of the tree if
the number of nodes $N$ is such that $N >$\ding{172}.
\end{corollary}

\textit{Proof.}  The  corollary   follows from
Theorems~\ref{t2},~\ref{t4_Turing}, and Corollary~
\ref{c4_Turing}. \hfill
 $\Box$

These results   lead to the following theorem again under the same
assumption about the chosen level of granularity of observations,
i.e.,   the nodes of the computational tree of $\mathcal{T^{IN}}$
representing configurations of the machine are observed.

\begin{theorem}\label{t5_Turing}
Given a non-deterministic Turing machine $\mathcal{T^{IN}}$ with a
depth, $k$,   of the computational tree and with a
non-deterministic degree $d$ such that
 \beq
 \frac{d(kd^{k+1}-(k+1)d^{k}+1)}{(d-1)^{2}}\leqslant\mbox{\ding{172}},
\label{Turing_9}
 \eeq
  then there exists an equivalent
deterministic Turing machine $\mathcal{T^{I}}$ which is able to
simulate $\mathcal{T^{IN}}$ and can be observed.
\end{theorem}

\textit{Proof.} For simulating $\mathcal{T^{IN}}$, the
deterministic machine $\mathcal{T^{I}}$ executes a breadth-first
visit of the computational tree of $\mathcal{T^{IN}}$. In this
computational tree, whose depth is
$1\leqslant{k}\leqslant$\ding{172}, each node has, by definition,
a number of children $c$ where $0\leqslant{c}\leqslant d$. Let us
suppose that the tree is complete, i.e., each node has $c=d$
children. In this case the tree has $d^{k}$ leaf nodes and $d^{j}$
computational paths of length $j$ for each $1\leqslant{j}\leqslant
k$. Thus, for simulating all the possible computations of
$\mathcal{T^{IN}}$, i.e., for a complete  visiting the
computational tree of $\mathcal{T^{IN}}$ and considering all the
possible computational paths consisting of $j$ computational steps
for each $1\leqslant{j}\leqslant k$, the deterministic machine
$\mathcal{T^{I}}$ will execute
 \beq
K_{\mathcal{T^{I}}}=\sum_{j=1}^k jd^j
 \label{Turing_14}
 \eeq
steps (note that if the computational tree of $\mathcal{T^{IN}}$
is not complete, $\mathcal{T^{I}}$ will execute less than
$K_{\mathcal{T^{I}}}$). Due to Theorems~\ref{t2}
and~\ref{t4_Turing}, and Corollary~ \ref{c5_Turing}, it follows
that in order to prove the theorem it is sufficient to show that
under conditions of the theorem it follows that
 \beq
K_{\mathcal{T^{I}}} \leqslant \mbox{\ding{172}}.
 \label{Turing_12}
 \eeq
To do this let us use the well known formula
 \beq
  \sum_{j=0}^k d^j =
\frac{d^{k+1}-1}{d-1}, \label{Turing_10}
 \eeq
 and derive both
parts of (\ref{Turing_10}) with respect to $d$. As the result we
obtain
 \beq
  \sum_{j=1}^k jd^{j-1} =
\frac{kd^{k+1}-(k+1)d^{k}+1}{(d-1)^{2}}. \label{Turing_11}
 \eeq
Notice now that by using (\ref{Turing_14}) it becomes possible to
represent  the number $K_{\mathcal{T^{I}}}$ as
\[
K_{\mathcal{T^{I}}}= \sum_{j=1}^k jd^{j} = d\sum_{j=1}^k jd^{j-1}.
\]
This representation together with (\ref{Turing_11}) allow us to
write
 \beq K_{\mathcal{T^{I}}} =
\frac{d(kd^{k+1}-(k+1)d^{k}+1)}{(d-1)^{2}}
 \label{Turing_15}
 \eeq
Due to assumption (\ref{Turing_9}), it follows that
(\ref{Turing_12}) holds. This fact concludes the proof of  the
theorem.
 \hfill  $\Box$

\begin{corollary}\label{c6_Turing}
Suppose that the length of the input sequence of symbols  of a
non-deterministic Turing machine $\mathcal{T^{IN}}$ is equal to a
number $n$ and     $\mathcal{T^{IN}}$ has a complete computational
tree with the  depth  $k$   such that $k=n^{l}$, i.e.,
polynomially depends on the length $n$. Then, if the values $d,
n$, and $l$ satisfy the following condition
 \beq
 \frac{d(n^{l}d^{n^{l}+1}-(n^{l}+1)d^{n^{l}}+1)}{(d-1)^{2}}\leqslant\mbox{\ding{172}},
\label{Turing_13}
 \eeq
then: (i) there exists a deterministic Turing machine
$\mathcal{T^{I}}$ that can be observed and able to simulate
$\mathcal{T^{IN}}$; (ii) the number, $K_{\mathcal{T^{I}}}$, of
computational steps required to a deterministic Turing machine
$\mathcal{T^{I}}$ to simulate $\mathcal{T^{IN}}$ for reaching a
final configuration exponentially depends on  $n$.
\end{corollary}

\textit{Proof.} The first assertion follows immediately from
theorem \ref{t5_Turing}. Let us prove the  second assertion. Since
the computational tree of $\mathcal{T^{IN}}$ is complete and has
the depth $k$, the corresponding   deterministic Turing machine
$\mathcal{T^{I}}$ for simulating $\mathcal{T^{IN}}$ will execute
$K_{\mathcal{T^{I}}}$ steps where $K_{\mathcal{T^{I}}}$ is from
(\ref{Turing_12}). Since condition (\ref{Turing_13}) is satisfied
for $\mathcal{T^{IN}}$, we can    substitute $k=n^{l}$ in
(\ref{Turing_15}). As the result of this substitution and
(\ref{Turing_13}) we obtain that
 \beq
K_{\mathcal{T^{I}}}=
\frac{d(n^{l}d^{n^{l}+1}-(n^{l}+1)d^{n^{l}}+1)}{(d-1)^{2}}
\leqslant\mbox{\ding{172}}, \label{Turing_16}
 \eeq
i.e., the number  of computational steps required to the
deterministic Turing machine $\mathcal{T^{I}}$ to simulate the
non-deterministic Turing machine $\mathcal{T^{IN}}$ for reaching a
final configuration is $K_{\mathcal{T^{I}}}
\leqslant\mbox{\ding{172}}$ and this number exponentially depends
on the length of the sequence of symbols provided as input to
$\mathcal{T^{IN}}$. \hfill  $\Box$

Results described in this section show that the introduction of
the new mathematical language including grossone allows us to
perform a more subtle analysis with respect to traditional
languages and to introduce in the process of this analysis the
figure of the researcher using this language (more precisely, to
emphasize the presence of the researcher in the process of the
description of automatic computations). These results show that
there exist limitations for simulating non-deterministic Turing
machines by deterministic ones. These limitations can be viewed
now thanks to the possibility (given because of the introduction
of the new numeral \ding{172}) to observe final points of
sequential processes for both cases of finite and infinite
processes.

Theorems~\ref{t4_Turing}, \ref{t5_Turing}, and their corollaries
show that  the discovered limitations and relations between
deterministic and non-deterministic Turing machines have strong
links with   our mathematical abilities to describe automatic
computations and to construct models for such descriptions. Again,
as it was in the previous cases studied in this paper, there is no
contradiction with the traditional results because both approaches
give results that are correct with respect to the languages used
for the respective descriptions of automatic computations.

We conclude this section by the note that analogous results can be
obtained for  physical machines $\mathcal{T^P}$ and
$\mathcal{T^{PN}}$, as well. In the case of imaginary machines,
the possibility of observations was limited by the mathematical
languages. In the case of physical machines they are limited also
by technical factors (we remind again the   analogy: the
possibilities of observations of physicists are limited by their
instruments). In any given moment of time   the maximal number of
iterations, $K_{max}$, that can be executed by physical Turing
machines can be determined. It depends on the speed of the fastest
machine $\mathcal{T^P}$ available at the current level of
development of the humanity, on the capacity of its memory, on the
time available for simulating a non-deterministic machine, on the
numeral systems known to human beings, etc. Together with the
development of technology this number will increase but it will
remain finite and fixed in any given moment of time. As a result,
theorems presented in this section can be re-written for
$\mathcal{T^P}$ and $\mathcal{T^{PN}}$ by substituting grossone
with $K_{max}$ in them.

\section{Conclusion}
\label{s5_Turing}

  The problem of mathematical descriptions of
automatic computations (the concept of the  Turing machine has
been used as a model of a device executing such computations) has
been considered in this paper from several points of view. First,
the problem has been studied using a new methodology emphasizing
in a strong form the presence   in the process of the description
of automatic computations of   the researcher who describes a
computational device and its properties. The role of the
philosophical triad -- the researcher, the object of
investigation, and tools used to observe the object -- has been
emphasized in the study. A deep investigation has been performed
on the interrelations that arise between mechanical computations
themselves and their mathematical descriptions  when a human (the
researcher) starts to describe a Turing machine (the object of the
study) by different mathematical languages (the instruments of
investigation).

 Along with traditional mathematical languages using such concepts as
`enumerable sets' and `continuum' to describe the potential of
automatic computations, a language introduced recently and the
corresponding computational methodology  allowing one to measure
the number of elements of different infinite sets have been used
in this paper. It has been emphasized that mathematical
descriptions obtained by using different languages depict the
object of the study -- the Turing machine~-- in different ways. It
has been established that the obtained descriptions, even though
in certain cases they give different answers to the same
questions, do not contradict each other. All of them are correct
with respect to the language used for the observation and the
description of the machines.

It has been established that there exists  the relativity of
mathematical descriptions of the object and there cannot be ever
any assurance that  a language chosen for the current description
expresses the object in an absolutely correct and complete way. A
richer language allows the researcher to reflect better the
properties of the studied object and a weaker language  does  this
worse (however, this fact can be noticed only if a richer language
is already known to the researcher). This situation is similar to
the work with a microscope where, when we need a more precise or a
more rough answer, we change the lens of our microscope. For
instance, suppose that by using a weak lens $A$ we see the object
of observation as one black dot while by using a stronger lens $B$
we see that the object of observation consists of two (smaller)
black dots. Thus, we have two different answers: (i) the object
consists of one dot; (ii) the object consists of two dots. Both
answers are correct with respect to the lens used for the
observation.

The new mathematical language applied in this study has allowed
the authors to establish a number of results regarding sequential
computations executed by the Turing machine and results regarding
  computable sequences produced by the machine. Deterministic and
non-deterministic machines have been studied using both the
traditional and the new languages. The obtained results have been
compared and discussed.

It has been emphasized that all mathematical (and not only
mathematical) languages (including the new one used in this study)
have   limited expressibilities. This fact leads to several
important reflections. First, for any fixed language there always
exist problems that cannot be formulated using it (these problems
often can be   seen when a new, sufficiently powerful for this
purpose language is invented). Second, there always exist problems
such that questions regarding these problems can be formulated in
a language but this language is too weak to express the desired
answer or the accuracy of the obtained answer is insufficient for
the practical needs. Finally, in any given moment of time for each
concrete problem there exists a finite number of languages that
can be used to attack the problem. Then, the most powerful
language among them defines computational bounds   for the problem
that exist both for physical and imaginary Turing machines (i.e.,
in both cases when a maximal finite or a maximal infinite number
of iterations is considered).




\bibliographystyle{plain}
\bibliography{Xbib_Turing}

\end{document}